# Online Misogyny Against Female Candidates in the 2022 Brazilian Elections: A Threat to Women's Political Representation?

Luise Koch[a]*, Raji Ghawi[a] Jürgen Pfeffer[a,b] and Janina Isabel Steinert[a,c]

[a]*TUM School of Social Sciences and Technology, Technical University Munich, Munich, Germany*,

[b]*School of Computer Science, Carnegie Mellon University, Pittsburg, United States of America*,

[c]*Department of Social Policy and Intervention, University of Oxford, Oxford OX1 2ER, United Kingdom*

**\*Corresponding author:** Luise Koch
**Email:** luise.koch@tum.de
**Telephone:** +49 175 686 2230
**Address:** Richard-Wagner-Straße 1, 80333 München

## Abstract

Technology-facilitated gender-based violence has become a global threat to women's political representation and democracy. Understanding how online hate affects its targets is thus paramount. We analyse 10 million tweets directed at female candidates in the Brazilian election in 2022 and examine their reactions to online misogyny. Using a self-trained machine learning classifier to detect Portuguese misogynistic tweets and a quantitative analysis of the candidates' tweeting behaviour, we investigate how the number of misogynistic attacks received alters the online activity of the female candidates. We find that young and left-wing candidates and candidates with higher visibility online received significantly more attacks. Furthermore, we find that an increase in misogynistic attacks in the previous week is associated with a decrease in female candidates' tweets in the following week. This potentially threatens their equal participation in public opinion building and silences women's voices in political discourse.

# Introduction

Women's participation in politics is not only fundamental to democracy but has been shown to be decisive for engendering policy outputs that benefit society as a whole. Female engagement in political decision-making is associated with improved well-being and health for women and children[1], enhanced governmental efficiency due to women-friendly or responsive policymaking[2,3] and higher levels of food security[4]. Female politicians also act as role models, thus raising the educational aspirations of girls[5] and ensuring democratic representation more generally[6].

Although women's participation and representation in executive, legislative and local decision-making bodies has increased over recent decades, progress towards equal representation remains slow. Only 26 countries worldwide are led by a female head of state or a female head of government, and only 26.5 per cent of all members of parliaments worldwide are female[7]. Hence, gender inequalities persist, and although women make up over half of the world population, they are still under-represented in political positions.

With the widespread use of social media platforms and the internet, political discourse has increasingly moved from offline spaces, such as town halls or rallies, to the digital realm. While this transition from offline to online communication has allowed politicians to establish closer links and more direct communication with their constituents[8], it has also become a breeding ground for hate speech[9]. This poses a threat to democracy as it can bias opinion-building processes, fuel populism, and shape political discourse in harmful ways[10].

Existing studies offer inconclusive findings, with some researchers providing evidence that female politicians receive more abuse[11,12] while others report the opposite[13,14]. However, while male politicians are more likely to be attacked because of their political views, female politicians are more likely to be harassed because they are women[15,16,17]. Thus, the online abuse that female politicians receive often takes the form of gendered and sexualised attacks on their appearance or character that objectify, belittle, and disrespect them[18,19]. This gendered form of abuse, referred to as *online misogyny*[20,21], can be considered a reaction to women entering a previously male-dominated political arena[22,23]. Against this backdrop, our study seeks to understand how female politicians are affected by online misogyny to assess its political implications.

A growing body of research has focused on the psychological effects on female politicians subjected to online abuse, using qualitative methods such as in-depth interviews and quantitative (online) surveys[24,25]. Analysing responses from an online survey, Akhtar et al. (2019) found that among 181 members of parliament in the UK, online hate had a much greater impact on female politicians, with women feeling more worried, fearful and concerned about their personal safety[26]. In Sweden, Erikson et al. (2021) conducted a survey and complementary semi-structured interviews, finding that female members of parliament were more likely to report that their freedom of expression online was circumscribed as a result of the online hate they received[27]. In keeping with this, more female than male members of parliament reported that they had reduced the frequency of their social media posts or had practised forms of self-censorship to avoid offensive comments and threats. Similarly, using in-depth interviews with male and female Canadian politicians who either had run, considered running or refused to run for office, Wagner (2022) found that gendertrolling deterred some female politicians from entering or staying in politics and contributed to their general perception of the work environment as hostile[28]. She also found that gendertrolling can have a silencing effect on women, with several interviewees reporting that they refrained from expressing feminist views online for fear of hostile backlash. Daniele et al. (2023) present the first causal evidence that violence targeted at politicians is driven by gender. Using data on offline and online violence against politicians in Italy over 12 years, the authors find that female politicians who were comparable to male politicians along sixteen observable characteristics received, on average, three times more attacks. The authors also found that attacked women were significantly less likely to remain in politics as compared to attacked men[29]. Taken together, previous evidence suggests that online hate may constitute a significant threat to women's representation in politics.

Building on this prior research, our paper aims to assess whether online misogyny has the potential to silence female politicians during campaign periods. Specifically, we present a large-scale longitudinal quantitative analysis of misogynistic attacks against 445 female candidates in the 2022 Brazilian election. Drawing on over 10 million tweets sent to or by these candidates, we examine how exposure to online misogyny affects candidates' online activity and, thus, their engagement in crucial political discourse and advocacy in the run-up to elections.

Our paper makes several important contributions. First, we argue that election campaigns are a particularly crucial period to examine as different candidates on the political spectrum compete for electoral success, providing fertile ground for online hate, intolerant rhetoric and incitement to violence[30,31]. Second, while efforts have been made to identify the toxicity of online hate more generally through artificial intelligence tools such as Google Perspective API, we develop and train a novel machine learning-based classifier that can detect the more nuanced nature of misogynistic Portuguese language. Third, we characterise female candidates according to their socio-demographic profile, political affiliation and online activity to systematically assess which candidates receive the most online hate. Last, while the majority of previous research draws on data from high-income countries, we shift the focus to Brazil, the Latin American country with the largest population and online audience and the world's fifth-largest social media user. Brazil has become highly polarised under former populist president Jair Bolsonaro. His political legacy and misogynistic rhetoric[32] have contributed to the persistence of patriarchal gender norms in Brazilian civil society[33], unprecedented levels of machismo and stereotypical gender representations in the media[34,35]. This climate has contributed to women's disempowerment and marginalisation from politics[36,37,] and has created a particularly hostile environment for those who have stood for election[38,39]. The Brazilian election of 2022, therefore, offers a particularly apposite crucible for assessing the vulnerability of female candidates to online misogyny and its political consequences.

# Results

To identify the number of misogynistic attacks, their targets and consequences for Brazilian female candidates, we developed a methodology that allowed us to capture the misogynistic attacks received by each candidate between January and November 2022. Using machine learning, we developed a classifier of misogyny and an analysis of the candidates' own tweeting behaviour.

First, our results indicate that female candidates were increasingly exposed to misogynistic attacks in the run-up to the elections. Second, we find that candidates with higher visibility online, young candidates and candidates with left to far-left political orientations received significantly more attacks. Third, we find evidence that misogyny can partly silence the voices of female candidates during election campaigns. Specifically, it is clear that if candidates received a large number of misogynistic attacks in a given week, the volume of their tweets declined in the following week relative to their average tweeting activity over time.

## Detecting Misogynistic Tweets

To capture the extent to which female candidates were targeted by misogynistic attacks on Twitter, we developed a machine-learning classifier for the Portuguese language. To this end, we drew on our dataset of 10 million tweets mentioning the 445 self-declared female candidates in the 2022 Brazilian election who had a Twitter account between January 1 and November 30. To train the classifier, we used a random subsample of 6000 tweets, which two coders coded manually as either misogynistic or not, achieving an inter-coder agreement of 0.785.

Applying the misogyny classifier to the full sample of more than 10 million tweets, 263,900 tweets (2.7%) were identified as misogynistic. Table 1 shows examples of tweets that were identified as misogynistic by our classifier, ranging from comments that objectify or infantilise candidates to more extreme threats of violence.

*Table 1:* Examples of misogynistic tweets

| Category | Examples | |
|---|---|---|
| | **English** | **Portuguese** |
| Body shaming | @[user] […] PS Joice, have you **put on weight again? You need to update the photo!** | @[user] […] PS Joice, vc engordou de Novo? Precisa atualizar a foto! |
| | @[user] […] **You're more than blonde.. the stupidity is ingrained..** | @[user] […] Vc é mais do que loira né.. a burrice tá enraizada... |
| | @[user] […] How you **like to lick a ball**, Joice you bootlicker, **relax my daughter**, your time will come, you too will be purged from politics. | @[user] […] Eita como gosta de lmaber umas bola heim Joice babaovo, relaxa minha filha sua hora vai chegar vc tbm vai ser expurgada da política. |
| Sexualisation/Objectification | @[user] As well as **being beautiful and sexy, you're intelligent.** I don't know how you still manage to remain faithful to the crackpot Jair [refers to Bolsonaro]. […] | @[user] Além de linda e sexy vc é inteligente. Não sei como vc ainda consegue permanecer fiel ao Jair. […] |
| Infantilisation | @[user] […] Ah! **Bye, darling.** You don't know anything about politics. Go back to journalism and start a blog! | @[user] […] Ah! Tchau querida. Você não entende nada de política. Volte para o jornalismo e crie um blog! |
| | @[user] […] **Stop being ridiculous** around here, don't you get tired **of embarrassing yourself**? Ask to leave for God's sake woman, **go do something useful in your life**! October is coming, your days are coming to an end…! | @[user] […] Pare de ser ridícula por aqui, vc não cansa de passar vergonha não? Pede pra sair pela amor de Deus mulher, vai fazer alguma coisa de útil na sua vida! Outubro tá chegando hein, seus dias estão acabando...! |
| Call for violence | @[user] […] **You deserve to be punished for so many lies. Punished by the law of men,** because the law of your God comes later. | @[user] […] A senhora merece ser punida por tantas mentiras. Punida pela lei dos homens, pq a do seu Deus vem depois. |
| | @[user] **What a disgusting woman**… But what can you expect? Even sewer rats aren't as filthy as you**.. If I walk past you, I'll throw shit at you**… | @[user] Que mulher nojenta… Mas esperar o quê de você né? Nem os ratos do esgoto são tão imundos como você... Se passo por você te jogo merda... |

*Note:* Translation for this collection of examples of misogynistic tweets was done by the authors. The name of the respective candidate was anonymised.

We also reveal that misogynistic attacks increased as the election date approached. Fig. 1 displays the absolute number of tweets mentioning the female candidates (left panel) and the percentage of misogynistic tweets (out of all tweets) per month (right panel). Figure S2 depicts the distribution of misogynistic mentions per candidate.

*Fig. 1:* Misogynistic mentions, own tweets and mentioned tweets over time

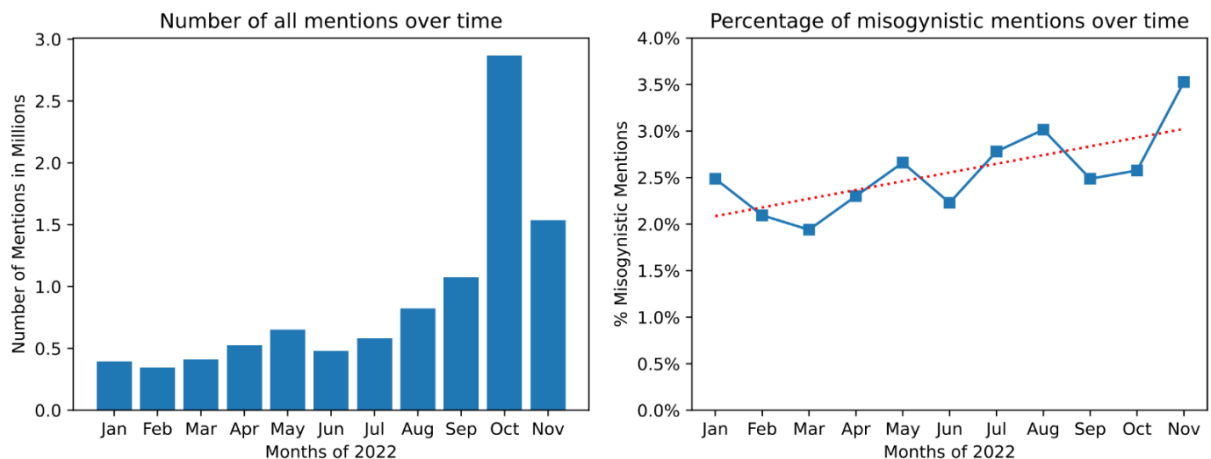

## Which Candidates Receive Most Misogynistic Attacks?

Following this, we sought to determine whether specific personal characteristics make a candidate more vulnerable to misogyny. For each of the 445 candidates that were mentioned more than 100 times, we collected information on ten personal characteristics, including their political orientation, ethnicity and age (see Table S2 in supplementary file). We then entered all characteristics simultaneously as covariates in a generalised linear poisson regression. The outcome variable was defined as the absolute number of misogynistic mentions candidates received between January and November 2022. The results of the regression analysis are summarised in Table 2.

*Table 2:* Generalised Linear Model Regression Results

| Variable | Coef. | Std. Err. | 95% CI |
|---|---|---|---|
| log10(*NumberOfMentionsCount*) | 2.5473*** | 0.013 | (2.522, 2.572) |
| log10(*FollowerCount*) | 0.3644*** | 0.014 | (0.337, 0.392) |
| log10(*OwnTweetCount*) | 0.1976*** | 0.011 | (0.176, 0.220) |
| *Region (reference group: REST)* | | | |
|   São Paulo | 0.1481*** | 0.011 | (0.126, 0.171) |
| *Party (reference group: REST)* | | | |
|   Partido Liberal | 0.0790*** | 0.021 | (0.037, 0.121) |
|   Partido dos Trabalhadores | 0.1575*** | 0.018 | (0.122, 0.194) |
| *Political Orientation (reference group: Center)* | | | |
|   Far-left | -0.7093*** | 0.025 | (-0.759, -0.660) |
|   Left | -1.0626*** | 0.013 | (-1.088, -1.037) |
|   Right | -0.7608*** | 0.015 | (-0.791, -0.731) |
|   Far-right | -1.1915*** | 0.019 | (-1.229, -1.154) |
| *Bolsonaro Supporter (reference group: No)* | | | |
|   Unknown | 0.3579*** | 0.016 | (0.327, 0.388) |
|   Yes | -0.4183*** | 0.021 | (-0.460, -0.377) |
| *Religion (reference group: Christian/Catholic)* | | | |
|   Afro-Brazilian | 0.0415 | 0.059 | (-0.075, 0.157) |
|   Evangelical | 0.4450*** | 0.019 | (0.408, 0.482) |
|   Other | -0.1124*** | 0.021 | (-0.154, -0.071) |
| *Ethnicity (reference group: White)* | | | |
|   Asian | 0.2454*** | 0.027 | (0.192, 0.299) |
|   Black | -0.0027 | 0.021 | (-0.045, 0.039) |
|   Brown | 0.2478*** | 0.020 | (0.209, 0.286) |
|   Indigenous | -1.0893*** | 0.057 | (-1.201, -0.978) |
| *Hierarchy Level (reference group: State deputy)* | | | |
|   President | -0.0019 | 0.030 | (-0.060, 0.056) |
|   Vice-president | -0.6131*** | 0.038 | (-0.707, -0.556) |
|   Governor | -0.5264*** | 0.029 | (-0.583, -0.470) |
|   Vice-governor | -0.2131** | 0.072 | (-0.354, -0.072) |
|   Senator | 0.3183*** | 0.023 | (0.273, 0.364) |
|   Federal deputy | 0.2814*** | 0.017 | (0.247, 0.315) |
| *Age Group (reference group: <+=30)* | | | |
|   Age Group: 31-40 | -0.2239*** | 0.018 | (-0.259, -0.189) |
|   Age Group: 41-50 | 0.0746*** | 0.017 | (0.042, 0.108) |
|   Age Group: 51-60 | -0.0717** | 0.020 | (-0.111, -0.032) |
|   Age Group: 61-70 | -0.2494*** | 0.016 | (-0.281, -0.218) |
|   Age Group: >70 | -0.6707*** | 0.030 | (-0.729, -0.613) |
| Constant | -7.6140*** | 0.048 | (-7.708, -7.520) |

*Note:* The table displays the coefficients, standard errors, p-values, and 95% confidence intervals for a generalized linear model regression with the Poisson family and log link function with 445 observations. The (Absolute) Number of misogynistic mentions is the **dependent variable**. The **Pseudo R-squ.(CS)** is 1.000 and **Covariance Type** is nonrobust. Significance levels are denoted as *** $p < 0.001$, ** $p < 0.01$, * $p < 0.05$.

First, it can be seen that candidates with higher visibility online were more vulnerable to misogynistic attacks. Specifically, we find that a tenfold increase in the number of mentions of a candidate (one-unit increase in the log10) was associated with a significant 2.5 increase in the absolute number of misogynistic attacks. Similarly, misogynistic attacks increase significantly with the number of followers on Twitter and with the number of tweets a candidate sends.

Apart from this, we find that candidates from the State of São Paulo received significantly more misogynistic mentions compared to candidates from other states. While candidates from Lula da Silva's party (Partido dos Trabalhadores) and Jair Bolsonaro's party (Partido Liberal) received significantly more misogynistic mentions than those from other parties, candidates from Lula's party received more misogynistic mentions than those supporting Bolsonaro's party. In terms of candidates' political orientation, we found that generally, positions left and right of the centre received fewer attacks, and the far right received the least. In line with this, candidates who openly declared themselves as Jair Bolsonaro supporters were also less likely to receive misogynistic attacks. Furthermore, candidates who identified as Evangelical received significantly more misogynistic mentions compared to candidates with other Christian or Catholic religious orientations. Candidates' minority status played a more ambiguous role. While Black candidates and candidates with indigenous backgrounds received fewer attacks than their White counterparts, Brown and Asian candidates faced significantly higher levels of misogyny relative to White candidates. Compared to the lowest hierarchy level of the state deputy, candidates running for higher positions, such as for the vice president, governor or vice-governor were targeted significantly less. Candidates running for senator and federal positions were slightly more targeted than state deputies. Lastly, we find that higher age correlated with fewer misogynistic attacks. Specifically, candidates in the age group below 30 years received significantly more misogynistic mentions than all other age groups and the number of misogynistic mentions was negatively correlated with increasing age (see Fig. S3 for boxplots visualizing the rate of misogynistic attacks for each category of the heterogeneity variables summarised in Table S2).

## How do Candidates' React to Misogynistic Attacks?

To understand candidates' reactions to misogynistic attacks, we analysed their tweeting behaviour on a weekly level. Specifically, we hypothesised that the online activity of a candidate would decrease after receiving misogynistic attacks online. To analyse candidates' reactions, we aggregated the daily number of misogynistic mentions and the daily number of their tweets per week. This served as a starting point for capturing potential variation in the baseline activity levels of the candidates, as some candidates may have habitually tweeted more than others. We applied a hierarchical clustering outlier detection algorithm to the weekly data points of each candidate to identify each candidate's baseline activity based on the inlier cluster and deviations from that activity based on the outliers. The outliers captured two types of possible behaviours: either less activity due to more misogynistic attacks or more activity associated with less misogynistic attacks.

As a result of our analysis, 1219 candidate-week pairs (19.5%) were identified as outliers showing behaviour diverging from the baseline activity. The average number of outliers per candidate was 8.5. Fig. 2 illustrates the inliers and outliers per week for two exemplary candidates, where each point represents the relationship between the number of misogynistic mentions in the previous week and the number of the candidate's tweets in the following week.

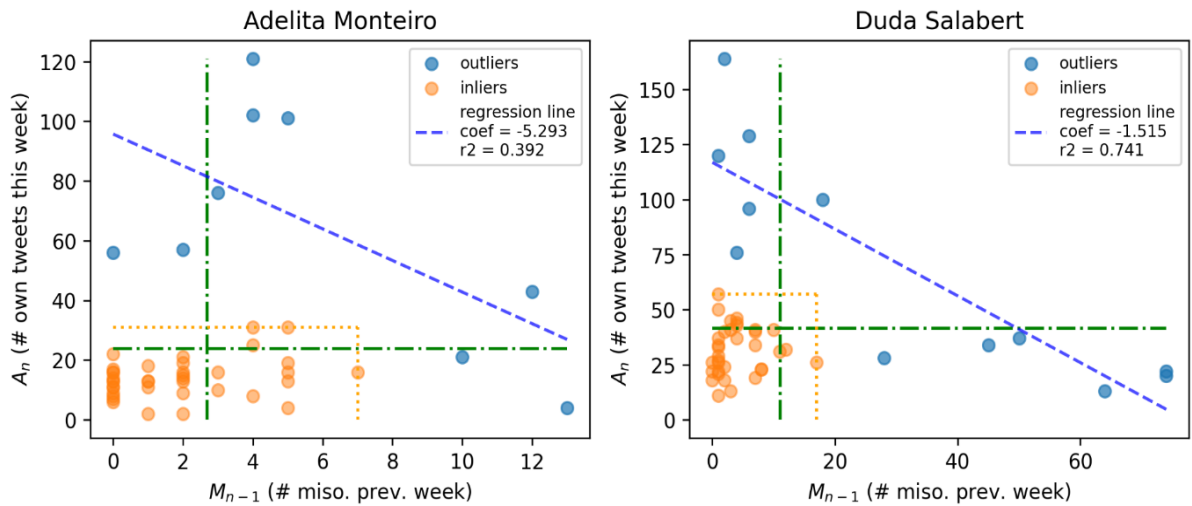

*Fig. 2:* Inlier and outlier plots for two exemplary candidate cases

*Note:* The dotted green lines depict the mean of the respective inlier and outlier groups.

The two figures demonstrate that most of the points are located around and below the orange cluster of inlier points, representing candidates' baseline activity level. The outlier points have been fitted to a regression line, which exhibits a negative slope, suggesting that more misogynistic mentions in a respective week correspond to less online activity in the following week. Mirroring this pattern, we can also observe that fewer misogynistic mentions in a respective week correspond to more tweeting activity in the following week. We fitted similar regressions for each candidate, showing that regression coefficients were negative for 94% of all candidates.

Subsequently, we aggregated the standardised outlier points across all candidates to estimate the overall deterrence effect as shown in Figure 3. (see Fig. S4 for the visualisation of both aggregated inliers and outliers). The cloud of outlier points is scattered around four quadrants created by the 0/0 lines. The lower right area shows that more misogynistic mentions in a respective week led to reduced activity in the following week. Furthermore, the upper left area shows that elevated activity in a given week is dominantly connected to below-average misogynistic mentions in the previous week. Lastly, the few observations in the upper right corner of the plot imply that more misogynistic mentions in a respective week were seldomly related to more activity in the following week.

*Fig. 3:* Cloud of standardised outlier points from all candidates

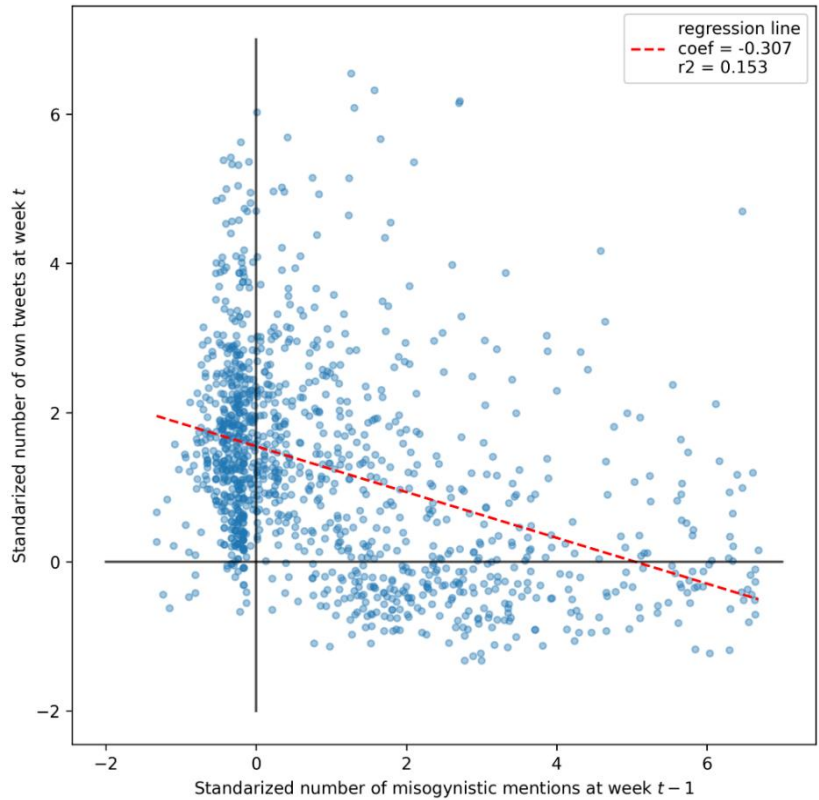

*Note:* Each point represents the relationship between the previous week's standardised number of misogynistic mentions and the candidate's tweets in the following week from all those candidates who received at least five misogynistic mentions per week (no. of candidates included=143).

The fitted regression line draws on the collective cloud of outlier points to estimate the relationship between the standardised number of tweets a candidate sent in a given week and the number of attacks she received in the prior week. The coefficient is negative and statistically significant, i.e. $\beta=-0.307$, p-value $< 0.001$ (see Table S3 for the full regression output). In other words, a one-unit standard deviation increase in the number of misogynistic attacks in the previous week led to a 0.3 unit standard deviation decrease in own tweets sent in the current week.

## Discussion

This study evaluated the extent and impact of online misogyny on female candidates in the Brazilian election of 2022. Applying a self-trained misogyny classifier to the full sample of 10 million tweets mentioning 445 self-declared female candidates in the 2022 election, we identified 263,900 tweets (2.7%) as misogynistic and observed an increase in misogynistic speech during the campaign period. Candidates were subject to more misogynistic attacks if they had greater visibility online, were younger, were left to far-left in political orientation, and were lower in their party hierarchy. Our study is also one of the first to provide quantitative evidence of a deterrence effect of misogynistic attacks on female politicians. Specifically, we showed that a one-unit standard deviation increase in misogynistic attacks received in the previous week correlated with a 0.3 standard deviation decrease in the volume of tweets that a candidate sent in the current week.

Our study builds on prior research in several ways. First, we move beyond the identification of online hate by studying the more subtle and difficult-to-detect phenomenon of online misogyny. Other tools, such as the Google Perspective API, have attempted to identify the extent of toxicity in certain posts; however, they are not well suited for detecting more subtle forms of misogynistic language in which words with inherently positive connotations may be used to objectify or sexualise women in a given context. To illustrate this, tweets that reduce female candidates to characteristics such as "beautiful", "faithful", or "elegant", without acknowledging their political competence, would be picked up by our misogyny classifier while scoring low on the Google Perspective API, thus demonstrating the more challenging endeavour of identifying misogynistic content due to its multiple layers and specific context dependence[40]. While projects such as the Automatic Misogyny Identification campaign have served as a starting point for the automatic detection of misogyny in English, Italian and Spanish[41,42], our classifier extends on these previous attempts by identifying misogyny in Portuguese. More importantly, our final stacked model outperforms previous misogyny classifiers in terms of accuracy and F1-measure (see Table S4 for a classifier comparison).

Second, our findings provide important insights into the extent of online misogyny that female candidates are exposed to during campaign periods –a particularly critical stage in the political cycle. We reveal that nearly 3 out of every 100 tweets received by female candidates in the Brazilian election were misogynistic in nature. While Agarwal et al. (2021) and Ward et al. (2017) observed comparable levels of hateful and abusive tweets received by members of the UK parliament outside of election periods (1% of all tweets and 2.6% of all tweets, respectively), our findings are striking in that they illustrate the dynamic nature of online misogyny. Specifically, we document a notable increase in the proportion of misogynistic tweets as the election approaches, suggesting that misogyny may serve as a strategic tool to undermine and weaken female politicians.

Third, our study adds novel context-specific insights into the intersectionality of online misogyny with characteristics such as age and ethnicity. Our heterogeneity analysis reveals that young and white candidates and candidates opposing the presidential candidate Jair Bolsonaro faced a disproportionate share of the misogynistic attacks during the 2022 elections in Brazil. In particular, our observation that white candidates were targeted more frequently than black and Indigenous candidates contrasts with a previous study from the UK that collected data about Members of Parliament in 2019, which found that black or minority candidates were targeted significantly more often[43]. This may be traceable to the particular socio-political context in Brazil but while we did not observe this pattern in our sample of 445 female candidates, it can still be assumed that individual black or minority candidates in Brazil are exposed to high levels of political and virtual violence[44,45]. In particular, the assassination of Rio de Janeiro city councillor Marielle Franco, a black lesbian politician and social justice activist by two police officers in 2018 highlights the high risks faced by women of colour who become politically active in Brazil. Apart from this, in line with Daniele et al. (2023), who found that female mayors were attacked regardless of their ideological orientation, our analysis revealed the same pattern for Brazilian state deputies. Compared to the lowest hierarchy level of state deputy, candidates running for higher positions, such as for vice president, governor or vice governor, were targeted significantly less, which is particularly worrisome as candidates running for local positions may be less able to protect themselves against the attacks received due to less professional support, financing or party backing.

Fourth, by drawing on more than 10 million tweets and automated detection of misogynistic language, we conduct a large-scale longitudinal analysis of misogynistic attacks and their impacts on female politicians during critical campaign periods. To this end, we drew on previous scholarship that aimed at exploring the deterrence effect of online hate. For example, Gorrell et al. (2020) found a positive, statistically significant relationship between being subjected to online abuse and the decisions of both male and female candidates to stand down during the six-week campaign period in the 2019 UK. The authors conclude that the harassers' aim of silencing politicians appears to have been effective. Similarly, Daniele et al. (2023) document that female and male mayors were generally equally likely to run for political office, but when they focused on the subsample of mayors who had received online and offline attacks, women were substantially more likely to leave politics in response to these. The authors characterise these attacks as a form of backlash against women's empowerment and political representation. Lastly, qualitative data collected from female politicians in Canada, the UK and the US suggests that although targets of online hate may refuse to adapt or give in to the attacks, they often believe that such attacks are part and parcel of modern politics and thus have to be endured[46,47,48].

Our study has some limitations. First, extremely hateful tweets may have already been removed by the platform, thus eluding our data capture. Relatedly, the identification of misogynistic tweets is a challenging task because attackers tend to disguise offensive words by inserting asterisks and spaces, replacing characters with similar sounds, or using abbreviations. In addition, the common use of irony, as well as the context-dependent and very subtle and sometimes benevolent nature of misogynistic language, makes detection difficult[49]. Our detection algorithm, despite its high accuracy and F1-measure, may, therefore, not have been able to identify the totality of misogynistic tweets. Taken together, this implies that we are likely underestimating the true extent of online misogyny against female candidates as well as its silencing effect.

Second, our study focuses exclusively on online misogyny experienced through Twitter and not through other platforms such as Instagram, TikTok or Facebook, via email or even in offline settings.

Experiences of online misogyny on Twitter could also have spillovers to offline behaviour — for example, through fewer public appearances — that we are unable to capture. We are not able to capture this offline deterrence effect, which is another reason why we likely underestimate the full deterrence effect. However, as social media has become increasingly prominent for political opinion formation and campaign purposes, analysing dynamics on Twitter can be considered as an important starting point.

Third, we have no information on whether the misogynistic attacks were read by the candidates themselves, whether they were filtered by the candidates' social media teams or whether they were ignored altogether. It is, therefore, possible that for some candidates, their online activity in a given week cannot be interpreted as a direct and causal response to the attacks they received in the previous week.

Fourth, we did not consider male candidates in the Brazilian election and the extent and impact of the hate attacks that they may have received. We are, therefore, unable to draw any relative gender comparisons in terms of how many attacks male and female politicians have received and to what extent they are deterred by these.

Lastly and most importantly, some female politicians may have decided not to run and campaign for the elections as a reaction to the misogyny they had previously been exposed to or anticipated to receive. These candidates are not captured in our dataset and are also not represented in our estimate of the deterrence effect. There is indeed evidence that some women are pushed out of politics at an earlier stage, i.e. before an election campaign, as illustrated by two prominent examples: Manuela D'Ávila, a former congresswoman from Rio Grande do Sul, who openly declared that she would not run again in the 2022 election because of the constant attacks on her family and herself from Bolsanoro and his supporters[50]. Similarly, the black congresswoman Aurea Carolina announced her withdrawal from the campaign period of the 2022 election to focus on her physical and mental health after holding a political mandate in which she was subjected to countless sexist and racist attacks from her colleagues and via social networks[51]. We also observed that 30 of the candidates included in our dataset had deactivated, deleted, privatised or suspended their Twitter accounts after the beginning of the election campaign, and three candidates, although running for a position, had turned on their tweet protection so that these were only visible to their followers. This could be another indication of a silencing mechanism and a coping strategy that some candidates adopt to shield themselves from hateful attacks. Further research is needed to determine whether our findings hold for other platforms, which may have different communication norms, community structures and algorithmic settings, and whether they also hold for male candidates.

Despite these limitations, our findings offer crucial new insights into how online misogyny can harm female politicians. First, we found that online misogyny is associated with reduced engagement and participation of women in political discourse during election campaigns, which may negatively impact their political campaigns and advocacy and ultimately lead to lower voter turnout. Second, experiencing and coping with online misogyny likely creates mental load and additional work for female candidates and their campaign teams[52], which consumes important resources that could otherwise be spent on campaigning. Third, being the target of online misogyny may prevent women from fully and equally participating in public and political life and may force some women to leave politics altogether. This not only negatively impacts their own campaigning but can also have a deterring effect on the next generation of female leaders who may be discouraged from entering politics altogether[53].

Taken together, our findings shed light on the situation of female candidates during election campaigns as political discussions have become ever more digitalised. Despite Brazil's 30 per cent quota for female candidates and all-time highs for female representatives and minority groups in the 2022 elections, online misogyny remains a critical burden and reflects persistent patriarchal patterns in Brazilian society[54]. Online misogyny against women politicians must be understood as a wider democratic challenge and a major threat to the UN Sustainable Development Goal 5, which aims to achieve gender equality and empower all women and girls. Online misogyny not only thwarts citizen-centred policy initiatives but also threatens to undermine social equality between male and female citizens.

# Methods

## Data Collection

X, formerly Twitter, was chosen as a suitable platform for this study as existing research shows that political elites and other actors have widely adopted the platform as a personalisation, mobilisation and promotion tool. As of 2022, Twitter was used by 18 million Brazilians and was one of the most used social media platforms for the country's online audience. In the run-up to the Brazilian election for president, congress, governors and legislatures of the 27 Brazilian states, we selected all self-declared female candidates from the official electoral tribunal side and checked whether they had a Twitter account (see Section 1.1 in the supplementary file for further detail on the sample selection). Of the 4292 female candidates running for election, we identified Twitter accounts for 977 candidates. This sample of candidates was further reduced to retain all 445 candidates who were mentioned at least 100 times between January 1 and November 30, 2022. We collected all tweets with academic API that mentioned any of the 445 candidates or were posted by them, which resulted in a total of 10,002,174 tweets (see Section 1.2 in the supplementary file).

## Classifier Training

To train the machine-learning classifier to detect misogyny in Brazilian Portuguese, we drew a random subsample of 6000 tweets from female candidates who were mentioned at least 100 times during the campaign period of the election (15.08.2022- to 02.10.2022). [Removed to maintain blinding] and a native speaker research assistant manually coded this sample of tweets into the binary categories of 'containing misogynistic language' and 'not containing misogynistic language'. In the coding process, 2.7% of the assessed tweets were identified as misogynistic, with an inter-coder agreement of 0.785 (the coding guidelines are specified in Section 1.3 in the supplementary file).

We used the coded dataset and applied a standard pre-processing of the tweet texts. Then, in order to restrict the vocabulary to the most significant words concerning misogyny, we opted for a keyness analysis[54] of each word concerning the misogyny variable. Subsequently, we split the dataset into 80% training and 20% testing subsets, using stratification to preserve the distribution of the misogynistic tweets within these subsets. The training subset was used to train a model, while the test subset was used for its validation. Section 1.4 of the technical appendix and Table S1 and Fig. S1 present an analysis of the different classification algorithms we tested to identify the best-performing algorithm regarding accuracy and F1-measure. Accuracy denotes the ratio of correctly predicted classifications of misogynistic vs. non-misogynistic language to the total number of instances and the F1-measure is defined as the harmonic mean of recall and precision of both classes. The classifier that best predicted the maximum possible number of misogynistic tweets was a stacking model combining five algorithms, achieving an overall accuracy of 91.1% and a F1-measure of 90.4%.

## Heterogeneity Analysis

To understand which candidates were more or less targeted by online misogyny, we assessed heterogeneity by the following characteristics: two continuously measured characteristics, including the number of Twitter followers and the number of own tweets sent by each candidate, and eight categorical characteristics, including the candidates' party hierarchy, political orientation, party affiliation, ethnicity, age, religion, region and whether the candidate openly declared support for Jair Bolsonaro or not. Table S2 reports the different categories and their grouping criteria. Each of the eight categorical variables was replaced by a set of binary variables per category. Of these, one binary variable was dropped and used as the reference category. To assess variation in exposure to online misogyny along the above-specified characteristics, we estimated a generalised linear poisson regression. The outcome variable was defined as the absolute number of misogynistic mentions received by each candidate between January and November 2022.

## Candidates' Reactions to Online Misogyny

Following the heterogeneity analysis, we analysed whether online misogyny had the potential to deter or silence its targets. We used the candidates' online activity as a proxy for their reactions to misogynistic attacks. First, we aggregated the number of misogynistic mentions of a respective candidate as well as the number of tweets posted on a weekly level. For each candidate $c$, we considered the relationship between the activity of candidate $c$ in week $t$, captured through the number of own tweets ($A_t^c$) and the misogynistic mentions received by candidate $c$ in week $t-1$ ($M_{t-1}^c$).

To capture variation in the baseline tweeting activity of the candidates, we detected any outliers in candidates' tweeting activity and thus deviations from their personally representative baseline behaviour. For outlier detection, we applied hierarchical clustering to the set of data points by using Euclidean distance as a distance measure between data points, and an average linkage as a distance measure between clusters. The goal was to identify one cluster of inliers and the rest as outliers. The inlier cluster was characterised by two requirements: First, the cluster had to be homogeneous, which we determined through the lowest average internal distance. Second, the cluster had to be located in the bottom left corner relative to other clusters, which meant that its centroid had to be less than the centroids of other clusters with respect to both dimensions, i.e. $M_{t-1}^c$ and $A_t^c$. In other words, the cluster had to have the lowest average both on the x-axis (misogynistic mentions received the previous week) and on the y-axis (activity this week). In most cases, the inlier cluster was the largest cluster. Independently of how many clusters we obtained, in all cases, we identified the same large homogenous bottom-left cluster of inliers based on the given characteristics. The remaining points were considered as outliers. Subsequently, we applied a linear regression model to the following form to the collective outlier points of all 143 candidates:

$$A_t = \beta_0 + \beta_1 M_{t-1} + \varepsilon \quad (1)$$

whereby $\beta_1$ captures any changes in candidates' online activity in week $t$ in relation to the misogynistic attacks received in week $t-1$.

# Acknowledgements

The authors are grateful for the input and advice received from two Brazilian experts, Maria Paula Russo Riva and Ladyane Souza and the support with the coding task from Marina Gaeta.
We further wish to thank Bavarian Research Institute for Digital Transformation (bidt) for funding this research study. All data needed to evaluate the conclusion in the paper are present in the paper and the Supplementary Materials. The code and data supporting this study are available via OSF: DOI 10.17605/OSF.IO/45GEH.

# SUPPLEMENTARY MATERIAL:

ONLINE MISOGYNY AGAINST FEMALE CANDIDATES IN THE 2022 BRAZILIAN ELECTIONS: A THREAT TO WOMEN'S POLITICAL REPRESENTATION?

*The technical appendix outlines how we trained and tested the machine learning algorithm to identify misogynistic language in the dataset scraped from the online platform X, formerly Twitter. It also provides details on the manual coding and the predictive performance of different tested models. The tables and figures show additional outcomes and robustness checks of both the heterogeneity and behavioural response analyses. The code and data supporting this study are available via OSF: DOI 10.17605/OSF.IO/45GEH.*

## CONTENTS

# Technical Appendix

## 1. Sample selection

- Based on the official Electoral Supreme Court site (see here: https://www.tse.jus.br/eleicoes/eleicoes-2022/eleicoes-2022), we included all names and ballot box names[1] for the female candidates running for president, governor, vice governor, senator or federal deputy positions once they were publicly announced on 15 August 2022.
- Between 15-31 August 2022, we searched for the Twitter accounts via the ballot box name as well as the actual names of these candidates on Twitter.
- If no Twitter account was found, we searched on Google, Instagram and Facebook with a fixed search term, i.e. first with the candidate's full name and then with the ballot box name to find a potential cross-reference to the candidates' Twitter account.
- From the total sample of 4292 self-declared female Brazilian candidates, 977 were found to have a Twitter account. A table of the 977 candidates with a Twitter account is available upon request to the authors.

## 2. Collection of Twitter data

- We collected Twitter data between 16 and 24 December 2022 using the Twitter API v2 via Twarc2 library for Python.
- Using Python programming language, we scraped all tweets either mentioning the candidates (excluding retweets) or sent by the candidates themselves[2] between 1 January to 30 November 2022 with one query per candidate per month, resulting in a total number of 10,747 queries.
- We found an unbalanced distribution of mentions, i.e., some candidates had very few mentions, and others had a lot of mentions; therefore, we opted to retain only candidates with at least 100 mentions, thus reducing our dataset to 445 candidates.
- The final number of collected tweets was 10,002,174, whereas some tweets mentioned more than one candidate. The number of unique tweets was 9,735,461.
- We proceeded with this set of tweets for the classifier training and the analysis.

## 3. Manual coding of tweets

- We randomly selected a subsample of 6109 tweets such that (1) each tweet was at least 100 characters long (excluding @mentions), (2) kept the over-time distribution in a way that we had observations from the whole time period, and (3) enforced balanced representation of the mentioned users. We additionally applied the Google Perspective API on the 6109 tweets to ensure that toxicity scores were uniformly distributed.
- The subsample of 6000 tweets was manually coded into the binary categories "misogynistic language" or "no misogynistic language". The coding task was performed based on the Portuguese tweets by a native speaker research assistant and Luise Koch.
- The binary coding task was conducted based on the definition of online misogyny adopted by Ging et al. (2019) and Massanari (2017): "Targeted harassment and abuse of women on the internet, mostly on social media platforms via abusive and sexist language or imagery as well as threats of violence". We further provided the research assistant with detailed coding instructions (see section 5 below), including examples of tweets that should be included or excluded.

---

[1] Candidates can choose a ballot name to run for an election, which might differ from their birth name; e.g.: real name: Hérika Siqueira Menezes Passos and ballot box name: Hérika da Virtuosa or real name: Ione Neves Cunha and ballot box name: Ione Brasil.

[2] Tweets sent by the candidates themselves were considered as our dependent variable and were only used in the behavioural response analysis of the paper.

- The research assistant started by first reading the typology and coding manual and then performed a test round of coding 50 exemplary messages (not part of the actual dataset) that had been previously annotated by LK. Upon completion of this training, the research assistant continued with coding the selected 6109 tweets.
- 19.6% of the 6109 tweets (i.e., 1200 tweets) were double-coded by the research assistant and Luise Koch. The inter-coder agreement was at 0.785, thus considered a "substantial agreement".
- Upon completion of the coding task, 821 of the 6109 tweets (13%) were categorised as misogynistic.
- Ethical considerations were taken into account for the training and supervision of the research assistant. Kennedy et al. (2022) highlighted the pressing concern that annotators may experience secondary trauma or other negative emotions, such as desensitisation, when annotating hate speech. We, therefore, provided the research assistant with Kennedy's suggested *Written Guide 7* to help her detect any possible changes in cognition and prevent symptoms of secondary trauma. The guideline advises the annotator to take breaks and not to imagine traumatic situations. The RA was asked to stay in close contact with Luise Koch if she noticed any symptoms of distress or experienced any negative emotions.

## 4. Development of the misogyny classifier

- We used a bag of words approach to translate the Tweet text into a quantifiable measure. First, we created a matrix containing the posts as rows and each word of the classified data as a column name. Second, we removed small tokens (length of 1 letter), @mentions, #hashtags, punctuation and non-alphabetical tokens (e.g., emojis). The tweets were all in Brazilian Portuguese; thus, the pre-processing steps were adapted to Portuguese.
- We opted for a keyness analysis of each word concerning the misogyny variable. Specifically, we applied a chi-squared test for each word to check the statistical significance of the word concerning its association with the misogyny variable (i.e. whether a tweet was classified as misogynistic or not). This step is useful to restrict the vocabulary to the most significant words (concerning misogyny), hence reducing its size and making it more focused. We retained only the words with a p-value < 0.20, resulting in ~ 500 words.
- For the training of the misogyny classifier, we split the coded dataset into 80% training and 20% testing subsets in a stratified way, preserving the distribution of the misogynistic tweets within these subsets. The training subset was used to train a model, while the test subset was used for its testing.
- We tested several classification algorithms, including Random Forests (RF), Logistic Regression (LR), k-Nearest Neighbors (KNN), Support Vector Machine with linear kernel (SVC_L), Naive Bayes with multinomial prior (NB_M) and with Gaussian prior (NB_G), as well as a BERT language model-based classifier (BERT) and a stacking model (Stack), which combines LR, RF, SVC_L, NB_M, and NB_G classifiers.
- Table S1 and Fig. S1 show the performance of these different classifiers in terms of true positives (TP), true negatives (TN), false positives (FP) and false negatives (FN), as well as accuracy (fraction of TP and TN to all cases), and F1-measure (harmonic mean of recall and precision) of both classes (0 class = non-misogyny, 1 class = misogyny), and the weighted average F1-measure. Fig. S1 depicts the classifiers in terms of TP and TN, where we can see a trade-off between these. For instance, while Naive Bayes classifiers give the best TP among others, their TN is low. On the other hand, KNN gives a very high TN but a very low TP.
- We made our final selection of the classifier based on the overall performance in terms of accuracy and the weighted average F1-measure, aiming to maximise both TP and TN. Accordingly, the best classifier is the stacking model (Stack), with an accuracy of 91.1%. Therefore, we adopt this model for the prediction of misogynistic tweets in the entire dataset.

## 5. Coding instructions

Categorising Tweets as Misogynistic or Non-Misogynistic: A Guideline

1. **Instructions**
   The population we investigate are female Brazilian candidates running for the upcoming elections. We have scraped tweets mentioning all the candidates with a Twitter account and have drawn a random sample of 6000 tweets. Each tweet has an ID number, and the name of the mentioned candidate was removed. The unit of coding is the tweet, and we would like you to code the tweets as either misogynistic (1) or non-misogynistic (0). Please read the full tweet before you make a coding decision. The respective tweets are listed line-by-line in column A, an Excel file, and your coding decision should be inserted in column B. In case of any doubts, please indicate your thoughts in the additional column C so as to discuss this with Luise Koch before a final coding decision is reached. Below, you can find the definition of what we consider misogyny and what not.

2. **Coding based on the "definition" of misogyny**
   Although several attempts have been made to define the concept of online misogyny, no universally accepted definition has been agreed upon. Its broadness and its interdisciplinarity character make the concept of misogyny difficult to grasp. The following definition attempts to capture the main character:

   *"Targeted harassment and abuse of women on the internet, mostly on social media platforms via abusive and sexist language or imagery as well as threats of violence."* Massanari (2017) and Ging et al. (2019)

   Yet, it becomes evident that the categorisation of comments is difficult in light of a **missing context**, i.e. without having detailed information on the perpetrator and the victim, on potential previous interactions and communication that they may have had, and on the contentious issues and topics that may feature in the attack**.** The coding approach should therefore focus on mentioned **adjectives** or **nouns** clearly attacking/ insulting/ ridiculing and criticising **women**.

3. Coding based on examples of misogynistic messages

| Key | Category | Detailed description | Examples (English) | | Exemplo (Portugues) | |
|---|---|---|---|---|---|---|
| 1 | **Body shaming/Ageism** | Unsolicited opinion stating or commenting about a target's body; its shape, size, appearance (Schlüter et al. 2021[1]) or its chronological age (Iversen et al. 2009[2]) | "You're too old to be here" | "[NAME] is unattractive both inside and out. I fully understand why her former husband left her for a man - he made a good decision." | *"Você está velho demais para estar aqui"* | *"[NAME] é pouco atraente, tanto por dentro como por fora. Eu entendo perfeitamente porque seu ex-marido a deixou por um homem - ele tomou uma boa decisão."* |
| 2 | **Sexualization/objectification** | Portraying and treating women as an object (Nussbaum 1995[3], Papadaki 2010[4]). Often occurring in the sexual realm defining women as worthy only in terms of the sexual pleasures and sexuality associated with their physical body (Vaes et al. 2010[5]) | "Fucking cocksucking cunt" | "You look so sexy in this dress" | *"Arrombada desgraçada do caralho"* | *"Você fica tão sexy com este vestido"* |
| 3 | **Infantilization** | Equating femininity or the actions of women with vulnerability, submission, and naivety thereby patronising and discrediting them (Carlson, 2012[6]) | "Kiddo, a bit of advice…" | "Calm down, woman. This just proves how unprepared you are." | *"Criança, um pouco de conselho..."* | *"Acalme-se, mulher. Isto só prova o quanto você está despreparada."* |
| 4 | **Attacks "ad feminem"** | Attacking some feature of a woman's character instead of the substance of the argument/ position itself, often in combination with stereotypes (derived from Sheng et al. 2021[7]) | "Go into the kitchen and make me a sandwich" | "You don't even know how to park your car, get out of here" | *"Vá para a cozinha e me faça um sanduíche"* | *"Você não sabe nem como estacionar seu carro, saia daqui"* |

| Key | Category | Detailed description | Examples (English) | | Exemplo (Portugues) | |
|---|---|---|---|---|---|---|
| 5 | **Call for violence including rape and death threats** | Reference/ threat to physical harm, sexual violence or reprisal (OHCHR 2019[8]), with no other thematic content | "Shut your mouth, you Communist journazist. Your time will come. Or do you think anyone can escape God's hands?" | "Bitch, I hope they burn your pussy!!!!" | *"Cala sua boca sua jornazista Comunista. Sua hora vai chegar. Ou vc acha que das mãos de Deus alguém pode escapar?"* | *"Gorda filha da puta, tomara q queimem a tua goiba!!!!"* |

*Note:* (1) The listed categories are not exclusive and might intersect

4. **Coding based on examples of toxic but non-misogynistic messages**

| | Idea | Detailed description | Examples (English) | | Exemplo (Portugues) | |
|---|---|---|---|---|---|---|
| 1 | **Any hate not specifically targeting women or specific characteristics denoted as "female"** | Bias-motivated, hostile, and malicious language targeted at someone (Siegel 2020) without the explicit attack or discrimination of women or characteristics accounted to be female | "Your idea of free kindergartens for everyone will just not work, you stupid" | "Spending and spending the tax money of the people.. you corrupt idiot " | *"Sua idéia de jardins de infância gratuitos para todos não será viável, sua estúpida"* | *"Gastar e gastar o dinheiro dos impostos do povo..sua corrupta"* |

## 6. Behavioural Response

- For the whole time period from January 1 to November 30, 2022, we relied on 47 weeks and 14.753 candidate-week pairs, which were reduced to 13.988 candidate-week pairs when excluding observations from the first week of January as we were missing a previous reference week for the first week. Given that the analysis required a minimum amount of misogynistic mentions per week per candidate, we excluded all candidates who had received less than five misogynistic mentions in any given week, thus yielding a final sample of 143 candidates and 6.253 candidate-week pairs.
- For the outlier detection, we explored different algorithms, including the local outlier factor, isolation forest, and elliptic envelope. Since none of these approaches provided plausible results from a visual perspective, we opted for a hierarchical clustering approach.

# Tables

*Table S1:* Performances of different classifiers

| Classifier | True Negatives | True Positives | False Negatives | False Positives | Accuracy | F1-measure class-0 (non-misogynistic) | F1-measure class-1 (misogynistic) | F1-measure Weighted Average |
|---|---|---|---|---|---|---|---|---|
| **Stack** | 1027 | 86 | 31 | 78 | 0.911 | 0.950 | 0.612 | 0.904c |
| **SVC_L** | 1018 | 90 | 40 | 74 | 0.907 | 0.947 | 0.612 | 0.902 |
| **NB_M** | 981 | 112 | 77 | 52 | 0.894 | 0.938 | 0.635 | 0.898 |
| **Bert** | 1025 | 79 | 33 | 85 | 0.903 | 0.946 | 0.572 | 0.895 |
| **LR** | 1033 | 70 | 25 | 94 | 0.903 | 0.946 | 0.541 | 0.891 |
| **NB_G** | 963 | 108 | 95 | 56 | 0.876 | 0.927 | 0.589 | 0.882 |
| **RF** | 1040 | 52 | 18 | 112 | 0.894 | 0.941 | 0.444 | 0.875 |
| **KNN** | 1052 | 3 | 6 | 161 | 0.863 | 0.926 | 0.035 | 0.807 |

*Table S2:* Heterogeneity variables

| Characteristic | Groups | | Grouping Criteria |
|---|---|---|---|
| Twitter Follower Count [degree of being known] | Numeric variable | | |
| Own tweet Count [capturing overall online activity] | Numeric variable | | |
| Hierarchy Level | 1 | President | Information was taken from the official page of the Superior Electoral Court (TSE). |
| | 2 | Vice-President | |
| | 3 | Senator | |
| | 4 | Governor | |
| | 5 | Vice-Governor | |

| Characteristic | Groups | | Grouping Criteria |
|---|---|---|---|
| | 6 | Federal Deputy | |
| | 7 | State Deputy | |
| Political Orientation | 1 | Far-left | Information was based on detailed background research on candidates' political stances, either based on their self-declaration or information drawn from grey literature and assessments of Brazilian research consultants. |
| | 2 | Left | |
| | 3 | Centre | |
| | 4 | Right | |
| | 5 | Far-right | |
| Candidates Party | 1 | PSOL | Far-left | Information was taken from the official page of the Superior Electoral Court (TSE) during the electoral campaign period. Parties were grouped into the five parts of the political spectrum based on the combination of self-declaration of parties and substantial research of the (grey) literature. For the heterogeneity analysis, we chose the top ten parties based on the number of representatives in our sample of 445 candidates, and the remaining parties were grouped as "REST". |
| | 2 | UP | Far-left | |
| | 3 | PSTU | Far-left | |
| | 4 | PDT | Left | |
| | 5 | PSB | Left | |
| | 6 | PT | Left | |
| | 7 | PCdoB | Left | |
| | 8 | PCB | Left | |
| | 9 | REDE | Left | |
| | 10 | PV | Left | |
| | 11 | MDB | Center | |
| | 12 | PSDB | Center | |
| | 13 | Podemos | Center | |
| | 14 | Solidaridade | Center | |
| | 15 | PSD | Center | |
| | 16 | Cidadania | Center | |
| | 17 | PROS | Center | |
| | 18 | Avante | Center | |
| | 19 | PMN | Center | |
| | 20 | AGIR | Center | |
| | 21 | Republicanos | Right | |
| | 22 | União Brasil | Right | |
| | 23 | PMB | Right | |
| | 24 | PP | Right | |
| | 25 | PTB | Right | |
| | 26 | PRTB | Right | |
| | 27 | Patriota | Right | |
| | 28 | PL | Far-right | |
| | 29 | Novo | Far-right | |
| | 30 | PSC | Far-right | |
| | 31 | PSL | Far-right | |
| Ethnicity | 1 | Indigenous | Information was taken from the official page of the Superior Electoral Court (TSE) during the electoral campaign period, where the candidates self-declared their ethnicity. |
| | 2 | Asian | |
| | 3 | White | |
| | 4 | Brown (Parda) | |
| | 5 | Black | |
| Age | Group 1: | 18-29 | Information was taken from the official page of the Superior Electoral Court (TSE) during the electoral campaign period, where the candidates self-declared their age.<br><br>Further decision criteria were based on respective age restrictions for each political position in the Brazilian context. |
| | Group 2: | 30-39 | |
| | Group 3: | 40-49 | |
| | Group 4: | 50-59 | |
| | Group 5: | 60-69 | |
| | Group 6: | <70 | |
| Religion | 1 | Evangelical | Information was taken from grey literature research and assessments of Brazilian consultants. |
| | 2 | Afro-Brazilian | |
| | 3 | Christian/ Catholic | |
| | 4 | Jewish | |
| | 5 | No | |

| Characteristic | Groups | | Grouping Criteria |
|---|---|---|---|
| Jair Bolsonaro Supporter | 1 | Yes | Information was taken from grey literature research and assessments of Brazilian consultants. For the second part of the heterogeneity analysis, those candidates who were in the groups of "unknown" were excluded from the analysis. |
| | 2 | No | |
| | 3 | Unknown | |
| Region | 1 | Acre | Information was taken from the official page of the Superior Electoral Court (TSE) during the electoral campaign period on the position the candidate was running for in the respective state. |
| | 2 | Alagoas | |
| | 3 | Amapá | |
| | 4 | Amazonas | |
| | 5 | Bahia | For the heterogeneity analysis, we chose the top ten regions, based on the number of candidates and the rest of the regions were grouped into a REST group. |
| | 6 | Ceará | |
| | 7 | Distrito Federal | |
| | 8 | Espírito Santo | |
| | 9 | Goiás | |
| | 10 | Maranhão | |
| | 11 | Mato Grosso | |
| | 12 | Mato Grosso do Sul | |
| | 13 | Minas Gerais | |
| | 14 | Pará | |
| | 15 | Paraíba | |
| | 16 | Paraná | |
| | 17 | Pernambuco | |
| | 18 | Piauí | |
| | 19 | Rio de Janeiro | |
| | 20 | Rio Grande do Norte | |
| | 21 | Rio Grande do Sul | |
| | 22 | Rondônia | |
| | 23 | Roraima | |
| | 24 | Santa Catarina | |
| | 25 | São Paulo | |
| | 26 | Sergipe | |
| | 27 | Tocantins | |

*Table S3:* Regression output collective outlier cloud

| Variable | Coef. | Std. Err. | 95% CI |
|---|---|---|---|
| Number of misogynistic mentions in previous week | -0.3069*** | 0.021 | (-0.347, -0.266) |
| Constant | 1.5467*** | 0.043 | (1.463, 1.630) |

*Note:* The table displays the coefficient, standard errors, p-values, and 95% confidence intervals for a linear model regression with 143 observations. The Own Tweets this week is the dependent variable. The R-squ. is 0.153 and Covariance Type is non-robust. Significance levels are denoted as *** $p < 0.001$, ** $p < 0.01$, *$p < 0.05$.

*Table S4:* Misogyny Classifier Comparison

| Classifier | Sample | Accuracy | F1-measure |
|---|---|---|---|
| Stacked model used in this paper | 6.109 | 0.911 | 0.904 |
| Anzovino et al. (2018)[9] | 4.454 | 0.773 | 0.355 |
| Engish AMI EVALITA (BERT)[10] | 10.000 | 0.624 | 0.439 |
| Italian AMI EVALITA (SVM Linear Kernel)[10] | 10.0000 | 0.772 | 0.577 |
| English AMI IberEval (BERT)[10] | 3.9770 | 0.758 | 0.499 |
| Spanish AMI IberEval (SVM Linear Kernel)[10] | 4.138 | 0.734 | 0.469 |

# Figures

*Fig. S1:* Performance of different classifiers regarding true positives and true negatives

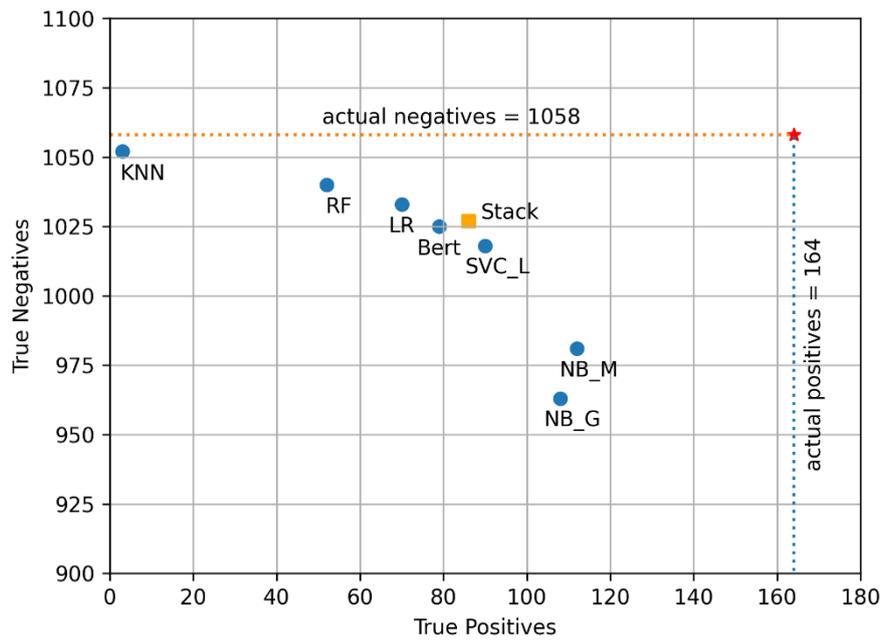

*Fig. S2*: Distribution of misogynistic mentions per candidate

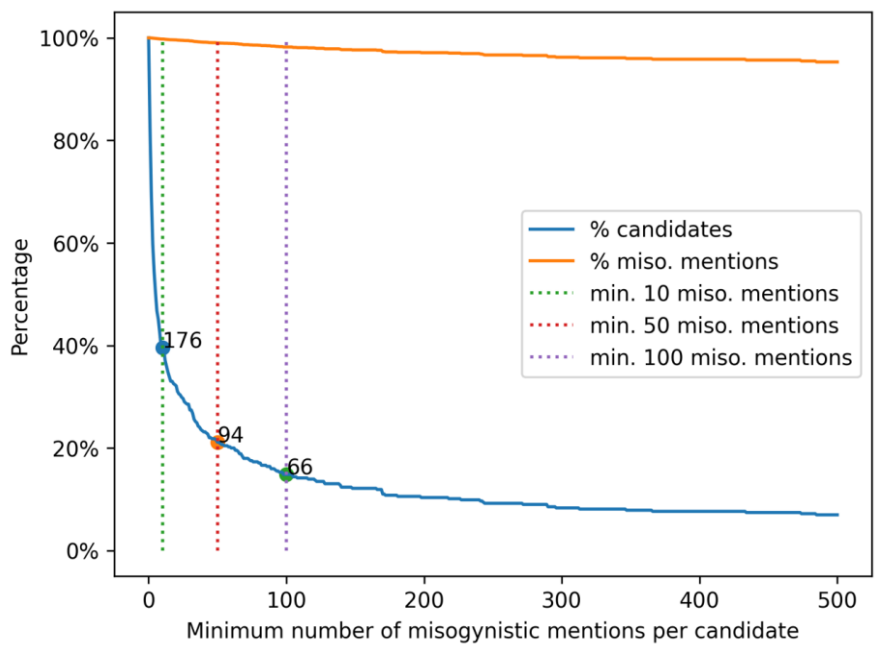

*Note:* 180 candidates received 10 or more misogynistic mentions (40%), 95 candidates received 50 or more mentions (21%), and 67 candidates received 100 or more misogynistic mentions (15%).

*Fig. S3*: Box plots of the relationship between misogyny received and heterogeneity characteristics.

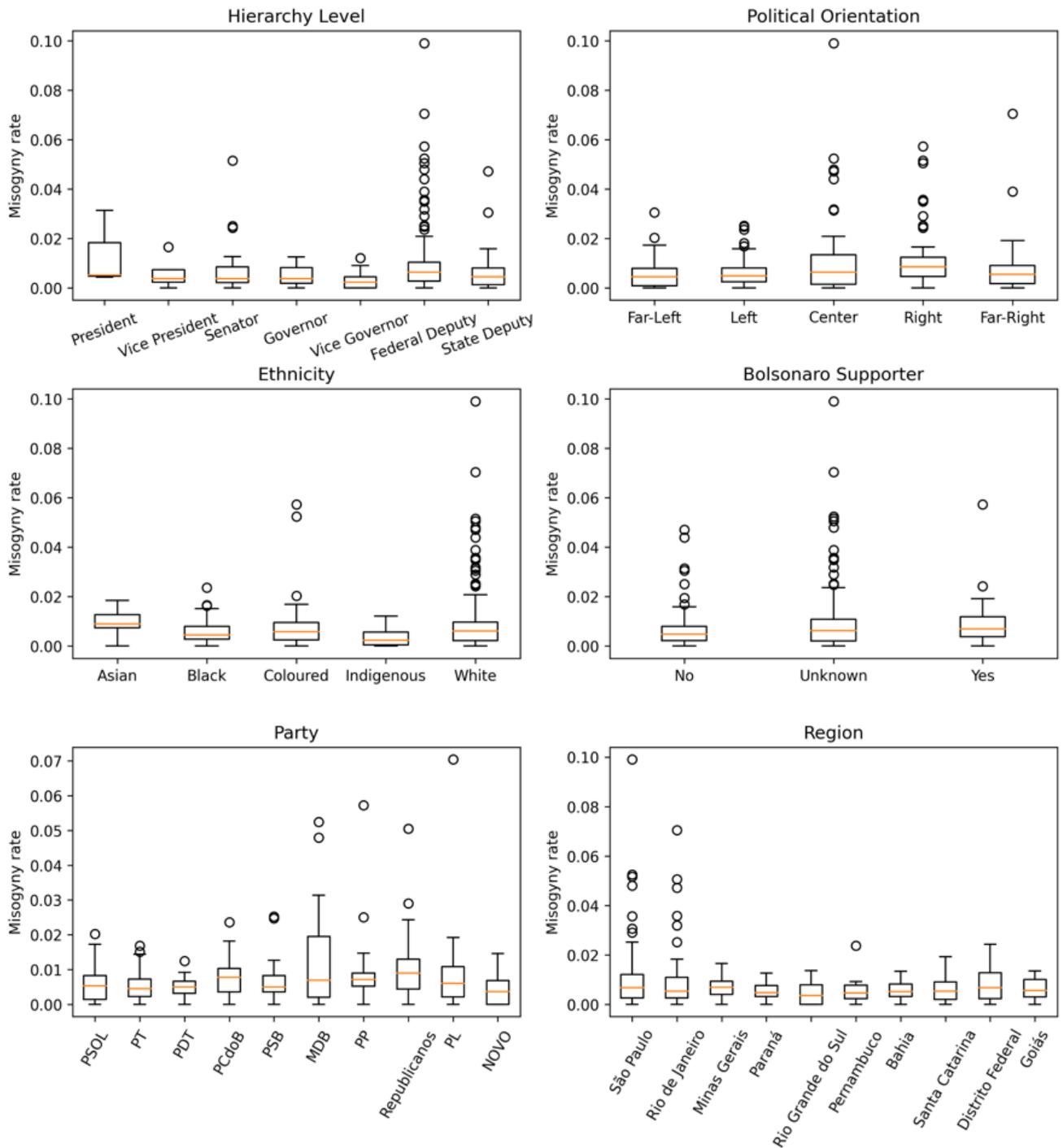

*Note:* The box plots visualise the misogyny rate per category of six heterogeneity variables and the prevalence of outliers. The ratio was calculated as the number of misogynistic mentions and overall mentions to obtain a relative measure of misogynistic mentions per candidate:

$$Ratio_{miso} = \frac{\#misogynistic\ tweets\ received}{\#overall\ tweets\ received}$$

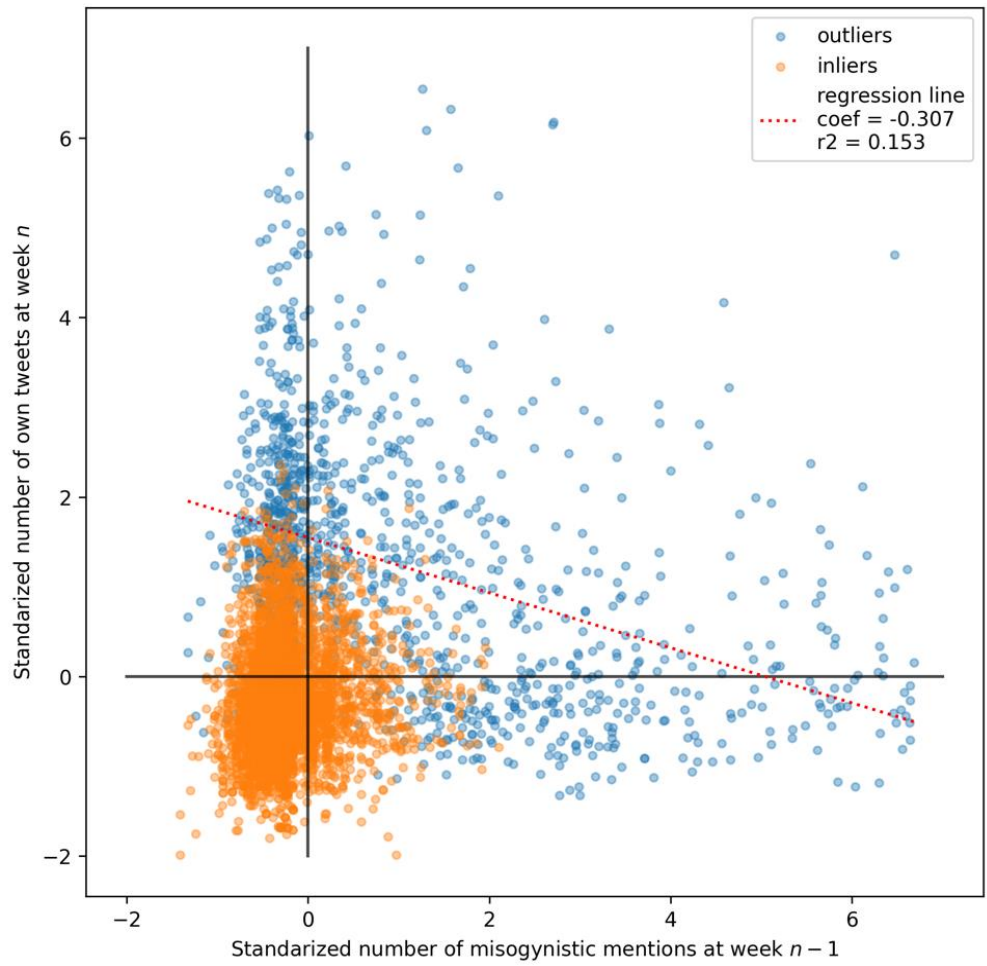

*Fig. S4:* Cloud of standardised outlier and inlier points from all candidates